%% file: main.tex
\documentclass[conference,comsoc,a4paper]{IEEEtran}
\IEEEoverridecommandlockouts

\usepackage{amsmath,amssymb,amsfonts}

\usepackage[ruled,norelsize]{algorithm2e}
\makeatletter
\newcommand{\removelatexerror}{\let\@latex@error\@gobble}
\makeatother
\SetKwInOut{Input}{Input}
\SetKwInOut{Output}{Output}

\usepackage{graphicx}
\usepackage{subcaption}
\usepackage{textcomp}
\usepackage{xcolor}
\usepackage[colorlinks=false, hidelinks]{hyperref}
\usepackage{cleveref}
\usepackage{siunitx}

\usepackage{tikz}
\usepackage{textcomp}
\usepackage{lipsum}

\usepackage[
backend=biber,
natbib=true,
url=false, 
doi=true,
eprint=false,
sorting=none,
maxbibnames=99,
defernumbers=true,
style=ieee,
giveninits=true,
]{biblatex}
\AtEveryBibitem{
 \clearlist{address}
 \clearfield{url}
 \clearfield{month}
 \clearfield{editor}
 \clearfield{eprint}
 \clearfield{volume}
 \clearfield{isbn}
 \clearfield{issn}
 \clearlist{location}
 \clearlist{editor}
}
\newcommand\copyrighttext{%
  \footnotesize \textcopyright 2024 IEEE. Personal use of this material is permitted.
  Permission from IEEE must be obtained for all other uses, in any current or future 
  media, including reprinting/republishing this material for advertising or promotional 
  purposes, creating new collective works, for resale or redistribution to servers or 
  lists, or reuse of any copyrighted component of this work in other works. 
  }
\newcommand\copyrightnotice{%
\begin{tikzpicture}[remember picture,overlay]
\node[anchor=south,yshift=10pt] at (current page.south) {\fbox{\parbox{\dimexpr\textwidth-\fboxsep-\fboxrule\relax}{\copyrighttext}}};
\end{tikzpicture}%
}

\usepackage{glossaries}

\def\BibTeX{{\rm B\kern-.05em{\sc i\kern-.025em b}\kern-.08em
    T\kern-.1667em\lower.7ex\hbox{E}\kern-.125emX}}

\input{my_macros}
\input{abbrevs}

\begin{document}

\title{Improving the Spatial Correlation Characteristics of Antenna Arrays using Linear Operators and Wide-band Modelling \\
\thanks{This work is partly sponsored by the \textit{Deutsche Forschungsgesellschaft (DFG)} under the research projects JCRS CoMP with Grant-No. TH $494/35-1$, HoPaDyn with Grant-No. TH $494/30-1$ and by the BMBF project 6G-
ICAS4Mobility with Project No. 16KISK241.} 
}

\author{
    \IEEEauthorblockN{
        Marc Miranda\IEEEauthorrefmark{1}, Sebastian Semper\IEEEauthorrefmark{1}, Michael Döbereiner\IEEEauthorrefmark{2}, Reiner Thomä\IEEEauthorrefmark{1}
    }\\
    \IEEEauthorblockA{\IEEEauthorrefmark{1} Technische Universit\"{a}t Ilmenau, Ilmenau, \{firstname.lastname\}@tu-ilmenau.de}
    \IEEEauthorblockA{\IEEEauthorrefmark{2} Fraunhofer IIS, Ilmenau, michael.doebereiner@iis.fraunhofer.de}
}

\maketitle

\copyrightnotice

\begin{abstract}
The analysis of wireless communication channels at the mmWave, sub-THz and THz bands gives rise to difficulties in the construction of antenna arrays due to the small maximum inter-element spacing constraints at these frequencies. Arrays with uniform spacing greater than half the wavelength for a certain carrier frequency exhibit aliasing side-lobes in the angular domain, prohibiting non-ambiguous estimates of a propagating wave-front's angle of arrival. 

In this paper, we present how wide-band modelling of the array response is useful in mitigating this spatial aliasing effect. This approach aims to reduce the grating lobes by exploiting the angle- and frequency-dependent phase-shifts observed in the response of the array to a planar wave-front travelling across it.

Furthermore, we propose a method by which the spatial correlation characteristics of an array operating at \carrier~GHz carrier frequency with an instantaneous bandwidth of \chaserbw~GHz can be improved such that the angular-domain side-lobes are reduced by 5-10~dB. This method, applicable to arbitrary antenna array manifolds, makes use of a linear operator that is applied to the base-band samples of the channel transfer function measured in space and frequency domains. By means of synthetically simulated arrays, we show that when operating with a bandwidth of \chaserbw~GHz, the use of a derived linear operator applied to the array output results in the spatial correlation characteristics approaching those of the array operating at a bandwidth of \targetbw~GHz. Hence, non-ambiguous angle estimates can be obtained in the field without the use of expensive high-bandwidth RF front-end components. 
\end{abstract}


\section{Introduction}
The estimation of the \gls*{aoa} of impinging wave-fronts serves as a key component in a number of existing applications such as radar, radio channel modelling, channel estimation in wireless communication networks and future applications such as \gls*{icas} \cite{van_trees_arrays_2002, krim_two_decades_1996, Xu_robust_doa_icas_2023}. To ensure non-ambiguous estimates of a narrow-band signal, it is required to sufficiently sample the wave-front in space, i.e. antenna inter-element distances must obey the spatial equivalent of the Nyquist theorem \cite{van_trees_arrays_2002}. The construction of antenna arrays that satisfy the maximum inter-element spacing requirement becomes increasingly difficult as we move towards the utilization of carrier frequencies in the \gls*{mmwave}, sub-\SI{}{\tera\hertz} and \SI{}{\tera\hertz} bands due to the increasingly small wavelengths. These challenges are mainly due to limitations in the fabrication process, difficulties in constructing feed networks for each element and the increased antenna mutual coupling that occurs at such small inter-element distances \cite{zihir_60ghz_2016}. Arrays that spatially \emph{under-sample} the impinging narrow-band wave-front cause aliasing in the angular domain and prevent non-ambiguous angle estimation unless additional information is made available to the estimator.

One method to mitigate these narrow-band aliasing effects is the exploitation of the frequency-dependent nature of wireless propagation, i.e. the use of excitation signals spanning a large enough bandwidth such that the narrow-band assumption is no longer valid. Since each propagating wave-front exhibits a frequency- and angle-dependent phase-shift \cite{weng_wideband_2023} when observed by the elements of the array, information about the angle of arrival can be obtained from the frequency domain samples of the channel. Methods exploiting frequency-dependent phase-shifts for joint \gls*{aoa} and \gls*{aod} estimation have recently been explored in literature, where the aim is to exploit the so-called reverse \emph{beam-squint} phenomenon \cite{luo_feifei_2023_beam_squint, boljanovic_cabric_2023_joint_mmwave_aoa_aod_one_ofdm}. These methods have been proposed considering novel architectures such as true-time-delay arrays that enable frequency-dependent beam-forming \cite{yan_cabric_2019_true_time_delay, boljan_yan_cabric_2021_fast_beam_true_time} at both transmitter and receiver.

In this publication, we propose a method by which the spatial correlation characteristics of an array operating at a certain bandwidth can be improved by making use of a linear operator to reduce the angular side-lobes that arise due large inter-element distances. This enables us to side-step the use of the large measurement bandwidth usually required to resolve the angle- and frequency-dependent phase shifts discussed earlier, hence reducing the cost of the RF components used in the radio front-end. We achieve this reduction by drawing on ideas from \gls*{cs} matrix design and re-purposing the approach taken by the authors in \cite{pawar_combining_2019, ibrahim_cmp_doa_2015}, where an improvement in spatial correlation characteristics was obtained for arrays with compressed outputs. 

The linear operator that achieves this improvement is derived numerically via a modified \gls*{sgd} algorithm for arbitrary array manifolds. While a model of the array manifold measured at a large bandwidth \emph{is} required during the design of the linear operator, channel measurements are carried out in the field at the \emph{lower} bandwidth. We show that estimation of the \gls*{aoa} of planar wave-fronts is possible even when the array inter-element spacing may exceed $\lamdaFactor \lambda$ and the bandwidth utilized is not large enough to resolve the individual wave-front phase-shifts (without further processing). Naturally, this approach is also applicable to the problem of  \gls*{aod} estimation when using antenna arrays at the transmitter. To model the frequency-dependent behaviour of the antenna array, we utilize the approach proposed by the authors in \cite{semper_high_2023} that extends the \gls*{eadf} \cite{landmann_efficient_2004} to antennas with extended \glspl*{ir} in the delay domain.

In summary, our proposed approach provides the following two advantages, namely

\begin{enumerate}
    \item enabling the use of antenna arrays with large inter-element spacing, hence easing the requirements on the fabrication process
    \item and subsequently enabling non-ambiguous angle estimates when utilizing the aforementioned arrays with bandwidths that would normally result in angular side-lobes, thus reducing the cost of RF components in the front-end.
\end{enumerate}

\subsection{Notation}
In this paper, we make extensive use of the Einstein notation to represent multi-way data arrays and tensors \cite{semper_high_2023}. Further, we use $(\cdot)^*$ to denote the complex-conjugate operation on an array. Using this notation, the outer product of two vectors is defined by aptly naming the sub-scripts used to label each axis, i.e.
\begin{equation*}
    \mathbf{C}_{p,q} = \mathbf{a}_{p} \cdot \mathbf{b}_{q} 
\end{equation*}
where $\mathbf{a}$ and $\mathbf{b}$ are vectors of arbitrary sizes. This notation allows us to simplify the representation of operations on multi-way arrays without having to represent and mainpulate them in vectorized form. To denote a single element of a multi-dimensional array $\mathbf{a}$, we use $\left[\mathbf{a}\right]_{pqr}$ where $p$, $q$ and $r$ denote the position of the element in each axis.

\section{Data Model}

We begin by considering an antenna array at the receiver with $N_R$ elements that samples an impinging wave-front in the far-field at $N_F$ frequency points spanning a total measurement bandwidth of $B$ \SI{}{\hertz}. The base-band samples of the measured channel transfer function for a single propagating path are therefore described by $\mathbf{s}: \left[ -\pi, \pi  \right) \times \R \rightarrow \C^{N_F \times N_R}$
\begin{equation}
    \mathbf{s}_{f,m}\left(\theta, \tau\right) = \gamma \cdot \mathbf{a}^{\text{rx}}_{f,m}(\theta) \cdot \mathbf{a}^{\text{f}}_{f}(\tau) + \mathbf{n}_{f,m},
    \label{eq:bb_samples}
\end{equation}
where $\gamma \in \C$ assumes a frequency-independent path weight, $\theta \in \left(-\pi,\pi\right]$ represents the azimuth angle of the impinging wave-front and $\tau \in \left(0,1\right]$ denotes the normalized path delay. The terms $\mathbf{a}^{\text{rx}}: (-\pi, \pi] \rightarrow \C^{N_F \times N_R}$ and $\mathbf{a}^{\text{f}}: (0,1] \rightarrow \C^{N_F}$ describe the polarimetric, far-field wide-band response of the array to impinging angle and delay, respectively. The indices $m$ and $f$ span the samples obtained in the spatial and frequency domain. While we consider only the azimuth angle here for ease of notation, the approach and results presented can be easily extended to consider full 2-D angle estimation.

If we assume the specular ray model of propagation \cite{richter_estimation_2005}, the received samples for $P$ propagating paths can be assumed to be a superposition of the responses of each path. Additionally, to signify the dependence of the antenna spatial response on the operating bandwidth, we introduce a new super-script as follows
\begin{equation}
    \mathbf{s}_{f,m} = \boldsymbol{\gamma}_p \cdot \mathbf{a}^{{\mathrm{B}},\text{rx}}_{f,m}(\boldsymbol{\theta}_p) \cdot \mathbf{a}^{\text{f}}_{f}(\boldsymbol{\tau}_p) + \mathbf{n}_{f,m},
    \label{eq:bb_samples_p_paths}
\end{equation}
where $\boldsymbol{\gamma} \in \C^P$, $\boldsymbol{\theta} \in \left(-\pi,\pi\right]^{P}$ and $\boldsymbol{\tau} \in \left(0,1\right]^P$.

The \gls*{scf} $\zeta: \R \times \R \rightarrow \C$ of the array operating at bandwidth $B$ \SI{}{\hertz} is defined as
\begin{equation}
    \zeta^{\mathrm{B}} \left( \theta_1, \theta_2 \right) = \mathbf{a}^{\mathrm{B}}_{m,f}(\theta_1)^* \cdot \mathbf{a}^{\mathrm{B}}_{m,f}(\theta_2),
    \label{eq:scf}
\end{equation}
and quantifies how strong the response of the array to a wave-front impinging from angle $\theta_1$ correlates to the response to angle $\theta_2$.

A common first step in \gls*{ml} based \gls*{aoa} estimation algorithms such as RIMAX \cite{richter_estimation_2005} is the computation of the correlation function $\mathcal{C}: (-\pi, \pi] \rightarrow \C$ defined as
\begin{equation}
    \mathcal{C}(\theta, \tau) = \mathbf{x}_{f,m}^{*} \cdot \mathbf{a}_{f,m} \left( \theta, \tau \right),
    \label{eq:corr_func}
\end{equation}
where $\mathbf{x} \in \C^{N_F \times N_R}$ represents a measured realization of the channel. A peak-search of this function provides estimators with an initial estimate of the non-linear parameters that is then further refined \cite{richter_estimation_2005}.

To critically sample a narrow-band planar wave-front in space, the spatial-Nyquist theorem states that the maximum inter-element spacing $d_{\text{max}} < \lambda_L/2$, where $\lambda_L$ is the wavelength corresponding to the highest frequency component used to excite the channel \cite{van_trees_arrays_2002}. The \gls*{scf} of an array violating this requirement exhibits strong grating lobes, as depicted in \Cref{fig:scf_patch_nb}. Similar grating lobes are observed in the correlation function \eqref{eq:corr_func} as illustrated in \Cref{sec:results}. 

In the next sections we describe how exploiting the frequency-dependency of the antenna response can result in a reduction in the magnitude of these grating lobes, as well as a method to improve the side-lobe suppression capabilities of the antenna array.

\section{Spatial Under-sampling and the narrow-band assumption}
The narrow-band propagation model as commonly used in angle and delay estimation algorithms such as ESPRIT \cite{wang_esprit_1989} or MUSIC \cite{schmidt_music_1986}, assumes that a wave-front undergoes only angle-dependent phase-shifts as it travels across the array, i.e., all frequency components impinging from a certain angle undergo identical phase-shifts \cite{richter_estimation_2005}. This assumption is valid when the the measurement bandwidth is small compared to the carrier frequency, giving rise to phase shifts (due to varying path lengths to each array element) that cannot be resolved for the given array geometry. The narrow-band assumption is particularly useful to reduce computational complexity in joint angle and delay estimation, as each non-linear path parameter (angle, delay) contributes to only one measurement domain (space, frequency) \cite{richter_estimation_2005}. 

\begin{figure}[h]
  \begin{subfigure}{\columnwidth}
    \centering
    \includegraphics[width=\linewidth]{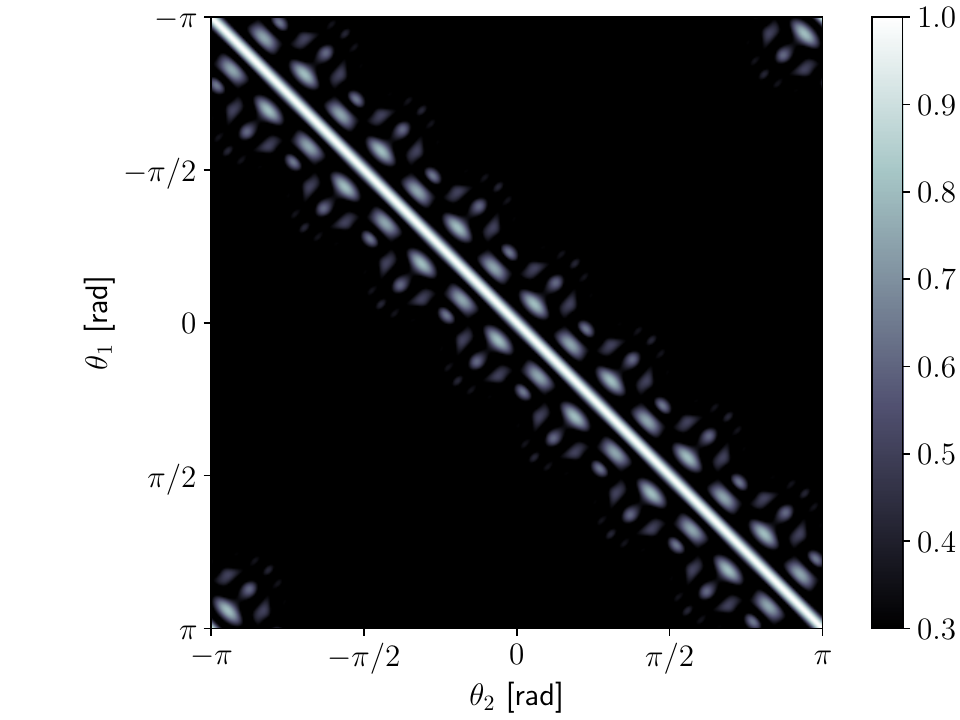}
    \caption{ $B_1 = \SI{\targetbw}{\giga\hertz}$  }
    \label{fig:scf_patch_wb}
  \end{subfigure}
  \begin{subfigure}{\columnwidth}
    \centering
    \includegraphics[width=\linewidth]{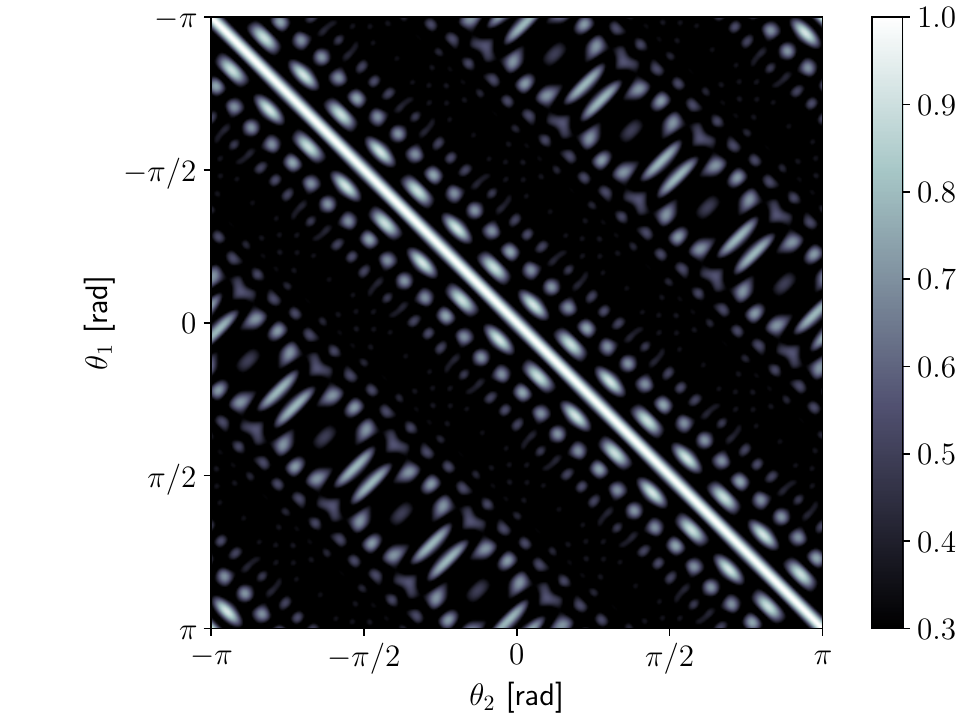}
    \caption{ $B_2 = \SI{\chaserbw}{\giga\hertz}$ }
    \label{fig:scf_patch_nb}
  \end{subfigure}
  \caption{Effect of measurement bandwidth on the magnitude of the \gls*{scf} $\left|\zeta(\theta_1, \theta_2)\right|$ for an array with \numElems patch elements arranged in a ring in the X-Y plane. The array is constructed such that it under-samples all impinging wave-fronts. The array considered in (b) operates at a frequency which can be considered narrow-band w.r.t the carrier frequency and exhibits strong off-diagonal elements corresponding to aliasing side-lobes while (a) is considered relatively wide-band and results in a significant reduction in angular-domain ambiguities.}
  \label{fig:scf_patch}
\end{figure}

Physically, however, every propagating wave-front exhibits different angle- and frequency-dependent phase-shift at each antenna element in the array \cite{weng_wideband_2023, boljanovic_cabric_2023_joint_mmwave_aoa_aod_one_ofdm}. By utilizing sufficiently large arrays and measurement bandwidths, the grating lobes observed in the \gls*{scf} of the array are suppressed by exploiting the individually resolvable phase-shifts. In \Cref{fig:scf_patch_wb}, we see how a large measurement bandwidth results in suppressed aliasing lobes by examining the \gls*{scf} of an array with inter-element spacing greater than $3\lambda_L$. Therefore, for a certain array geometry, an increased measurement bandwidth results in lower angular side-lobes.

A large bandwidth, however, correspondingly requires the use of expensive high-frequency RF components in the front-end. In this next section, we propose the design and utilization of a linear operator $\boldsymbol{\Phi}$ that improves the characteristics of the \gls*{scf} \eqref{eq:scf} for an array operating at a significantly smaller bandwidth, such that non-ambiguous estimates of the \gls*{aoa} can still be obtained from the measured data \eqref{eq:bb_samples}.

\section{Reducing Angular Side-lobes} \label{sec:reducing_angular_side}
\subsection{Linear Operator}
We consider two operating bandwidths of interest $B_1 \gg B_2$, such that a system employing $B_1$ \SI{}{\hertz} has significantly better spatial correlation characteristics than an identical system employing bandwidth $B_2$ \SI{}{\hertz}, as illustrated in \Cref{fig:scf_patch}. We define the tensor
\begin{equation}
    \boldsymbol{\Phi} \in \C^{M_F \times M_R \times N_F \times N_R},
\end{equation}
where $M_F$ and $M_R$ denote the array output size after applying the operator. The tensor $\boldsymbol{\Phi}$ operates on the samples of the \emph{lower} bandwidth system, i.e.
\begin{equation}
    \hat{\mathbf{s}}^{\mathrm{B_2}}_{f_1,m_1} 
        = \boldsymbol{\Phi}_{f_1,m_1, f_2,m_2} 
          \cdot \mathbf{s}^{\mathrm{B_2}}_{f_2,m_2}.
    \label{eq:extension_op_1}
\end{equation}
For brevity, we denote the indices of the input dimensions of $\boldsymbol{\Phi}$ by the vector quantity $\mathbf{m} = [f_2, m_2] \in \mathbb{N}^2$ and the output dimensions by $\mathbf{n} = [f_1, m_1] \in \mathbb{N}^2$, such that \eqref{eq:extension_op_1} is equivalently expressed as
\begin{equation}
    \hat{\mathbf{s}}^{\mathrm{B_2}}_{\mathbf{n}} = \boldsymbol{\Phi}_{\mathbf{n}, \mathbf{m}} \cdot \mathbf{s}^{\mathrm{B_2}}_{\mathbf{m}} \ \in \C^{M_F \times M_R}.
    \label{eq:extension_op_2}
\end{equation}

The \emph{effective} \gls*{scf} of a system employing this operator is described as
\begin{align}
    \hat{\zeta}^{\mathrm{B_2}} \left(\boldsymbol{\Phi}, \theta_1, \theta_2 \right) &= 
        \hat{\mathbf{a}}^{\mathrm{B_2}}_{\mathbf{n}}(\boldsymbol{\Phi}, \theta_1)^* 
        \cdot \hat{\mathbf{a}}^{\mathrm{B_2}}_{\mathbf{n}}(\boldsymbol{\Phi}, \theta_2) \nonumber \\
    &= \mathbf{a}^{\mathrm{B_2}}_{\mathbf{k}}(\theta_1)^* 
        \cdot \boldsymbol{\Phi}^*_{\mathbf{m}, \mathbf{k}} 
        \cdot \boldsymbol{\Phi}_{\mathbf{m}, \mathbf{n}} 
        \cdot \mathbf{a}^{\mathrm{B_2}}_{\mathbf{n}}(\theta_2),
    \label{eq:eff_scf}
\end{align}
where $\hat{\mathbf{a}}^{\mathrm{B_2}}: \C^{M_F \times M_R \times N_F \times N_R} \times \R \rightarrow \C^{M_F \times M_R}$ is the effective array response after using the linear operator. Computing the correlation function \eqref{eq:corr_func} for a certain measurement of the channel  $\mathbf{x} \in \C^{N_F \times N_R}$ with an array employing $\boldsymbol{\Phi}$ is similarly described by
\begin{equation}
    \mathcal{C} \left( \theta, \tau \right) = \mathbf{x}^{*}_{\mathbf{n}} \cdot \boldsymbol{\Phi}^{*}_{\mathbf{p}, \mathbf{n}} \cdot \boldsymbol{\Phi}_{\mathbf{p}, \mathbf{m}} \cdot \mathbf{a}^{B}_{\mathbf{m}}\left( \theta, \tau \right).
    \label{eq:corr_func_phi}
\end{equation}

The goal of using the operator $\boldsymbol{\Phi}$ is to improve the effective spatial characteristics in $\hat{\zeta}^{\mathrm{B_2}}$ such that they approach that of the system operating with bandwidth $B_1$ and characterized by $\zeta^{\mathrm{B_1}}$. For the sake of simplicity, we set the output dimensions of the operator to match the input, i.e. $M_F = N_F$ and $M_R = N_R$ since we are currently not interested in reducing the \emph{number} of measurement points in the spatial and frequency domain.

\subsection{Operator Design} \label{sec:extension_op_design}
While literature on \gls*{aoa} estimation using \gls*{cs} usually advocates for the use of sensing matrices (the linear operator $\boldsymbol{\Phi}$ in our formulation) with randomly drawn elements, previous work from our group shows that careful design of the matrices improves estimation performance \cite{pawar_combining_2019, lavrenko_combining_2018, ibrahim_cmp_doa_2015}.

To derive an effective realization of the operator $\boldsymbol{\Phi}$, we extend the approach outlined in \cite{pawar_combining_2019} to multiple measurement domains, i.e., space and frequency. We define an error metric $\delta: \C^{N_F \times N_R \times N_F \times N_R} \rightarrow \R$ that evaluates the improvement in spatial correlation characteristics obtained after utilizing the operator as
\begin{equation}
    \delta\left( \boldsymbol{\Phi}\right) 
        = \int\limits_{\theta_1} \int\limits_{\theta_2} 
            \left\vert e\left( \boldsymbol{\Phi}, \theta_1, \theta_2 \right)\right\vert^2 
            d\theta_1 d\theta_2,
    \label{eq:delta_phi}
\end{equation}
where $e: \C^{N_F \times N_R \times N_F \times N_R} \times \R^2 \rightarrow \C$ is defined as
\begin{align}
    e(\boldsymbol{\Phi},& \theta_1, \theta_2) = 
        \hat{\zeta}^{\mathrm{B_2}}(\boldsymbol{\Phi},\theta_1, \theta_2) 
        - \zeta^{\mathrm{B_1}}(\theta_1, \theta_2) \nonumber \\
    &= \hat{\mathbf{a}}^{\mathrm{B_2}}_{\mathbf{n}}(\boldsymbol{\Phi}, \theta_1)^* 
        \cdot \hat{\mathbf{a}}^{\mathrm{B_2}}_{\mathbf{n}}(\boldsymbol{\Phi},\theta_2) 
        -\mathbf{a}^{\mathrm{B_1}}_{\mathbf{n}}(\theta_1)^*
        \cdot \mathbf{a}^{\mathrm{B_1}}_{\mathbf{n}}(\theta_2).
    \label{eq:err_func}
\end{align}
The operator is then obtained via a minimization of $\delta\left( \boldsymbol{\Phi} \right)$
\begin{equation}
    \boldsymbol{\Phi}_{\mathrm{opt}} = \mathrm{arg} \min_{\boldsymbol{\Phi}} \delta(\boldsymbol{\Phi}),
    \label{eq:phi_opt_cont}
\end{equation}
for which a closed-form solution exists only for antenna arrays with row-orthogonal array manifolds \cite{ibrahim_cmp_doa_2015}. Since evaluating \eqref{eq:err_func} is unwieldy for 2-D angle estimation, we approximate the spherical integrals in \eqref{eq:delta_phi} by a discrete summation \cite{pawar_combining_2019}, such that minimizing
\begin{equation}
    \delta(\Phi) = \sum_{\theta_1} \sum_{\theta_2} \left|e(\boldsymbol{\Phi},\theta_1, \theta_2) \right|^2,
    \label{eq:delta_phi_discrete}
\end{equation}
leads to an equivalent solution for the problem in \eqref{eq:delta_phi} when the angles drawn on the unit sphere  are sufficiently dense and large in number.

\subsection{Gradient Based Optimization}\label{sec:grad_based_optim}
We now introduce the set of angles $\varphi^{(G)} = \{\theta_i | \theta_i \in \boldsymbol{\theta}^{(G)}, i = 1,2,\cdots,N_{\theta}\}$ drawn from the range $\boldsymbol{\theta}^{(G)} \in \left(\theta_L, \theta_H\right]$ where the parameters $\theta_L$ and $\theta_H$ allow for defining an angular region of interest. The error function \eqref{eq:err_func} evaluated on this grid $\varphi^{(G)}$ defined as $\mathbf{e}: \C^{N_F \times N_R \times N_F \times N_R} \rightarrow \C^{N_\theta \times N_{\theta}}$ is expressed as
\begin{equation}
    \mathbf{e}_{i,j}(\boldsymbol{\Phi}) = \mathbf{C}^*_{\mathbf{m},i}\cdot \boldsymbol{\Phi}^*_{\mathbf{p}, \mathbf{m}} \cdot  \boldsymbol{\Phi}_{\mathbf{p}, \mathbf{n}} \cdot \mathbf{C}_{\mathbf{n},j}
     - \mathbf{T}^*_{\mathbf{m}, i} \cdot \mathbf{T}_{\mathbf{m}, j}
    \label{eq:err_func_full_grid}
\end{equation}

where 
\begin{align}
    &\mathbf{T}_{\mathbf{m}, i} = \mathbf{a}^{B_1}\left(\varphi_i^{(G)}\right) \in \mathbb{C}^{N_F \times N_R \times N_{\theta}}, \nonumber \\
    &\mathbf{C}_{\mathbf{m}, i} = \mathbf{a}^{B_2}\left(\varphi_i^{(G)}\right) \in \mathbb{C}^{N_F \times N_R \times N_{\theta}} 
    \label{eq:T_and_C}
\end{align}
contain the response functions of the arrays operating at bandwidth $B_1$ and $B_2$, respectively, evaluated on the angles $\varphi^{(G)}$. To obtain an operator $\boldsymbol{\Phi}$ that minimizes the difference in \glspl*{scf} for these angles, we re-write \eqref{eq:phi_opt_cont} as
\begin{equation}
    \boldsymbol{\hat{\Phi}}_{\text{opt}} = \text{arg}\min_{\boldsymbol{\Phi}} E(\boldsymbol{\Phi}),
    \label{eq:phi_opt_final}
\end{equation}
where $E(\boldsymbol{\Phi}) = \mathbf{e}_{i,j}(\boldsymbol{\Phi})^* \cdot \mathbf{e}_{i,j}(\boldsymbol{\Phi}) \in \R$ represents the squared $\ell$-2 norm of \eqref{eq:err_func_full_grid}  and $\boldsymbol{\hat{\Phi}}_{\text{opt}} \rightarrow$ $\boldsymbol{\Phi}_{\text{opt}}$ for a large and dense grid of angles $\varphi^{(G)}$. 

The gradient of the scalar objective in \eqref{eq:phi_opt_final} w.r.t $\boldsymbol{\Phi}$, i.e. $\nabla_{\boldsymbol{\Phi}} E(\boldsymbol{\Phi}) \in \C^{N_F \times N_R \times N_F \times N_R}$ can be obtained via Wirtinger calculus \cite{pawar_combining_2019} as
\begin{align}
    \nabla_{\boldsymbol{\Phi}} &E(\boldsymbol{\Phi})_{\mathbf{p}, \mathbf{m}} = \nonumber \\ &4 \boldsymbol{\Phi}_{\mathbf{p}, \mathbf{d}} \cdot \mathbf{C}_{\mathbf{d}, i}  \cdot \mathbf{C}_{\mathbf{m}, i}^* \cdot \boldsymbol{\Phi}_{\mathbf{p}, \mathbf{m}}^* \cdot \boldsymbol{\Phi}_{\mathbf{p}, \mathbf{n}} \cdot \mathbf{C}_{\mathbf{n}, i} \cdot \mathbf{C}_{\mathbf{m}, i}^* \nonumber \\ &- 2 \boldsymbol{\Phi}_{\mathbf{p}, \mathbf{d}} \cdot \mathbf{C}_{\mathbf{d}, j} \cdot \mathbf{T}_{\mathbf{n}, j}^* \cdot \mathbf{T}_{\mathbf{n}, i} \cdot \mathbf{C}_{\mathbf{m}, i}^*,
    \label{eq:obj_derivative}
\end{align}
and enables the use of gradient-based optimization routines. In practice, the size and density of the grid $\varphi^{(G)}$ on which the functions \eqref{eq:T_and_C} are computed must be very large, such that a direct computation of \eqref{eq:obj_derivative} is unwieldly due to large memory and compute requirements. To improve computational tractability, we use the \gls*{sgd} algorithm \cite{pawar_combining_2019, Kingma_adam_2014} to split the optimization into steps requiring $K$ \emph{batches} of $S \ll N_{\theta}$ angles such that \eqref{eq:phi_opt_final} can be approximated by
\begin{equation}
\boldsymbol{\hat{\Phi}}_{\mathrm{opt}} \approx \arg \min_{\boldsymbol{\Phi}} \frac{1}{K} \sum_{k = 1}^{K} D^k (\boldsymbol{\Phi}),
	\label{eq:opt_problem_stochastic}
\end{equation}
where $D^k (\boldsymbol{\Phi}) = \mathbf{e}^k_{i,j}(\boldsymbol{\Phi})^* \cdot \mathbf{e}^k_{i,j}(\boldsymbol{\Phi}) \in \R$ is computed for the $k$-th batch of angles and 
\begin{equation}
	\mathbf{e}^k: \mathbb{C}^{N_F \times N_R \times N_F \times N_R} \rightarrow \mathbb{C}^{S \times S}
 \label{eq:err_func_batch}
\end{equation} 
is evaluated on a smaller grid size $S$. Note that now $\boldsymbol{\hat{\Phi}}_{\text{opt}} \rightarrow \boldsymbol{\Phi}_{\text{opt}}$ for large $K$ \cite{pawar_combining_2019}. The gradient \eqref{eq:obj_derivative} can similarly be approximated by
\begin{equation}
	[\nabla_{\boldsymbol{\Phi}} E(\boldsymbol{\Phi)}] \approx \frac{1}{K} \sum_{k = 1}^K \nabla_{\boldsymbol{\Phi}} D^k (\boldsymbol{\Phi}).
	\label{eq:gradient_stochastic}
\end{equation}

Furthermore, to improve convergence rate and reduce the sensitivity of the optimization steps on the learning rate $\alpha$ we make use of the ADAM \cite{Kingma_adam_2014} stochastic optimization routine such that adaptive learning rates are estimated for each element in $\boldsymbol{\Phi}$ during run-time. If we denote the gradient array computed at the $k$-th iteration $\mathbf{d}^k = \nabla_{\boldsymbol{\Phi}} D^k(\boldsymbol{\Phi}^k)$, the update of the momentum \cite{Kingma_adam_2014} array $\mathbf{z} \in \C^{N_F \times N_R \times N_F \times N_R}$ at each iteration follows
\begin{equation}
	\mathbf{z}^{k+1}  = \beta_1 \mathbf{z}^k + (1-\beta_1)\mathbf{d}^k,
\label{eq:momentum_update}
\end{equation}
while the squared-gradient accumulator \cite{Kingma_adam_2014} is similarly updated as
\begin{equation}
\mathbf{v}^{k+1}_{\mathbf{p}} = \beta_2 \mathbf{v}^k_{\mathbf{p}} + (1-\beta_2) \mathbf{d}^k_{\mathbf{p}}\cdot \left(\mathbf{d}^k_{\mathbf{p}}\right)^*,
\label{eq:accum_update}
\end{equation}
where $\beta_1, \beta_2 \in [0,1]$ denote the decay factors \cite{Kingma_adam_2014} and $\mathbf{v} \in \R^{N_F \times N_R \times N_F \times N_R}$. Note that in \eqref{eq:momentum_update}, \eqref{eq:accum_update} the super-script denotes the iteration index while the subscripts follow from the Einstein notation as before.

Finally, after initializing $\boldsymbol{\Phi}_{\text{opt}}$ with entries drawn from a standard complex normal distribution, we update the array at each iteration as
\begin{equation}
	\boldsymbol{\Phi}^{k+1}_{\mathbf{p}} = \boldsymbol{\Phi}_{\mathbf{p}}^k - \alpha \mathbf{z}^k_{\mathbf{p}} \cdot \mathbf{w}^k_{\mathbf{p}}
	\label{eq:phi_update}
\end{equation}
where the elements of $\mathbf{w}^k \in \R^{N_F \times N_R \times N_F \times N_R}$ expressed as 
\begin{equation}
    [\mathbf{w}^k]_{abcd} = \left( [\mathbf{v}^k]_{abcd} + \epsilon \right)^{-1/2}
\end{equation}
scale the step-size $\alpha$ for each element of $\boldsymbol{\Phi}$ and $\epsilon \ll 1$ accounts for possible divide-by-zero errors. The complete design algorithm is presented in \Cref{alg:phi_sgd}.

\begin{figure}[!t]
 \removelatexerror
\begin{algorithm}[H]
\DontPrintSemicolon
\Input{%
    \begin{tabular}[t]{ll}
      $K$ & No. of batches \\
      $N_{\mathrm{angles}}$ & Batch size \\
      $\theta_{\mathrm{L}}, \theta_{\mathrm{H}}$ & Angular Focus Region \\
    \end{tabular} \qquad
    \begin{tabular}[t]{ll}
      $\mathbf{a}^{B_1}$ & Low B/W Model \\
      $\mathbf{a}^{B_2}$ & Large B/W Model \\
      $\alpha$ & Step-size \\
      $\beta_1, \beta_2$ & Decay factors
    \end{tabular}
  }
\Output{$\boldsymbol{\hat{\Phi}}_{\text{opt}}$ Optimized Compression Operator}
$\boldsymbol{\Phi}_0 \gets \text{random initialization}$\;
$\mathbf{z}_0 \gets \mathbf{0}$\;
$\mathbf{v}_0 \gets \mathbf{0}$\;
$k \gets 0$\;
\While{$k < K$}{
    Draw random $\boldsymbol{\theta}^k \in (\theta_L, \theta_H)^{N_{\mathrm{angles}}}$\;
    Compute $\mathbf{d}^k$ on grid $\boldsymbol{\theta}^k$\;
    $\mathbf{z}^{k+1}  = \beta_1 \mathbf{z}^k + (1-\beta_1)\mathbf{d}^k$\;
    $\mathbf{v}^{k+1}_{\mathbf{p}} = \beta_2 \mathbf{v}^k_{\mathbf{p}} + (1-\beta_2) \mathbf{d}^k_{\mathbf{p}}\cdot \left(\mathbf{d}^k_{\mathbf{p}}\right)^*$\;
    $\boldsymbol{\Phi}^{k+1}_{\mathbf{p}} = \boldsymbol{\Phi}_{\mathbf{p}}^k - \alpha \mathbf{z}^k_{\mathbf{p}} \cdot \mathbf{w}^k_{\mathbf{p}}$\;
    $k = k + 1$\;
}
Return $\boldsymbol{\Phi}_k$
\caption{Algorithm to obtain $\boldsymbol{\hat{\Phi}}_{\text{opt}}$ using ADAM.}
\label{alg:phi_sgd}
\end{algorithm}
\end{figure}

\section{Numerical Verification}
\subsection{Synthetic Array Configuration}
To verify the proposed method, we consider a ring-shaped synthetic array with \numElems \ patch elements and inter-element spacing $d=\lamdaFactor\lambda_L$. The array is excited at \numFreq \ frequency points covering a bandwidth of $B_2 = \SI{\chaserbw}{\giga\hertz}$ centered around the carrier frequency of \SI{\carrier}{\giga\hertz}, with $\lambda_L$ corresponding to the wave-length of the lowest frequency sub-carrier. The far-field beam-pattern of each patch element is modelled as described in \cite{balanis_antenna_2005}. Further, in order to represent the frequency-selective response of the antenna array, we utilize the extensions to the \gls*{eadf} \cite{landmann_efficient_2004} presented in \cite{semper_high_2023} that model the slowly-varying frequency domain response efficiently by a small set of Fourier coefficients.

To re-iterate, we design the operator $\boldsymbol{\Phi}$ to be used in the field with the array operating at $B_2 = \SI{\chaserbw}{\giga\hertz}$, a bandwidth considered feasible for RF-circuitry currently available in channel sounding equipment\cite{muller_uwb_channel_sounder_2015}. However, we additionally require a simulated manifold of the same array for the larger bandwidth of $B_1 = \SI{\targetbw}{\giga\hertz}$ to design the operator following \Cref{sec:extension_op_design}. 

\begin{figure}[h]
    \centering
    \includegraphics[width=0.9\linewidth]{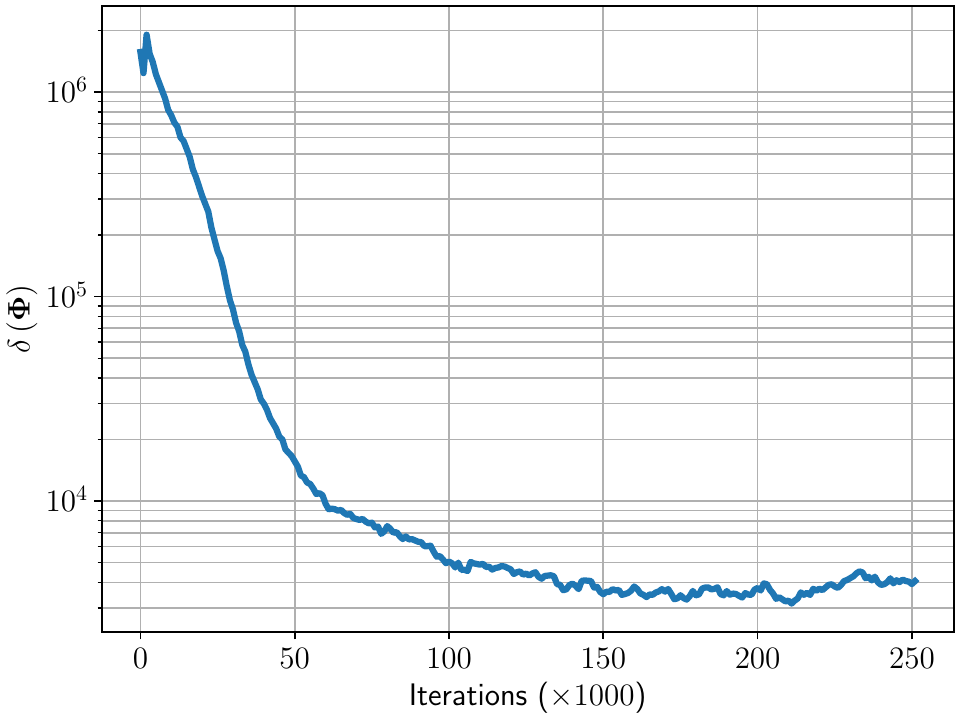}
    \caption{Evolution of the training error during stochastic optimization of the objective \eqref{eq:delta_phi}.}
    \label{fig:err_progress}
\end{figure}

\subsection{Stochastic Optimization}
As described in \Cref{sec:grad_based_optim}, we make use of the ADAM \cite{Kingma_adam_2014} stochastic optimization routine to minimize the objective function in \eqref{eq:delta_phi}. The algorithm is run with ADAM parameters (see \Cref{sec:grad_based_optim}) $K = 250\times 10^3$, $N_{angles} = 50$, $\beta_1 = \betaOne$, $\beta_2 = \betaTwo$, $\epsilon = 10^{-15}$ and $\alpha = \alphaLearn$. We set $(\theta_L,\theta_H] = (-\pi,\pi]$ such that random angles are drawn from the complete unit circle during the design routine. In \Cref{fig:err_progress} we see the progression of the magnitude of the error function \eqref{eq:err_func} computed over a fixed grid which is unrelated to the angular grid from which angles are drawn during optimization.

\begin{figure}[h!]
    \begin{subfigure}{\columnwidth}
    \centering
    \includegraphics[width=\linewidth]{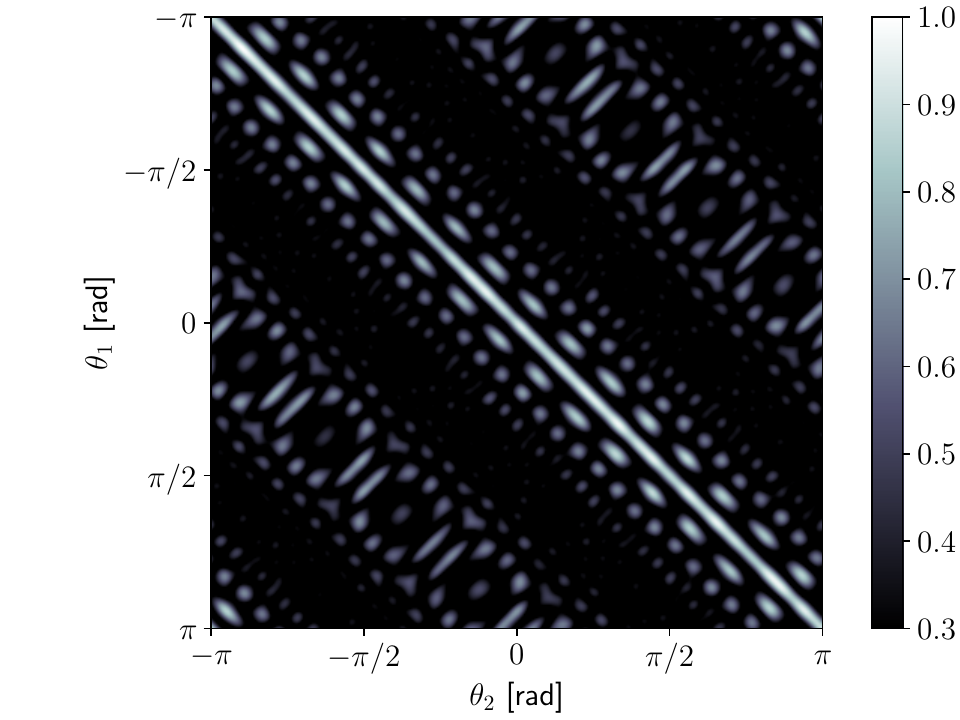}
    \caption{Random $\boldsymbol{\Phi}$}
    \label{fig:scf_patch_random}
  \end{subfigure}
  \begin{subfigure}{\columnwidth}
    \centering
    \includegraphics[width=\linewidth]{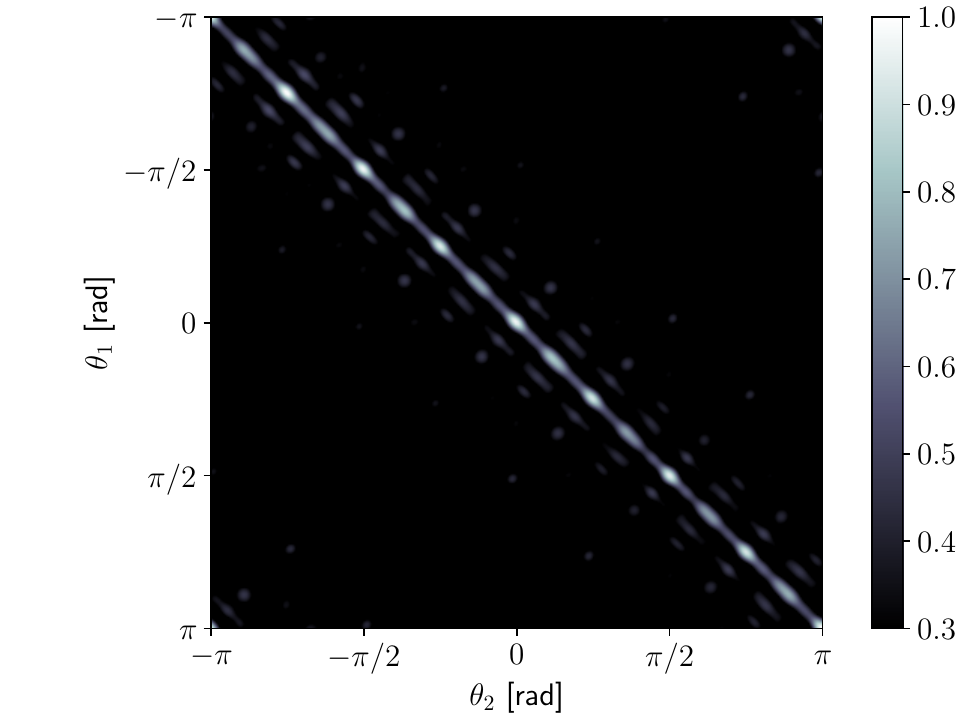}
    \caption{Optimized $\boldsymbol{\Phi}$}
    \label{fig:scf_patch_mixed}
  \end{subfigure}
    \caption{Magnitude of the effective \gls*{scf} $\left|\hat{\zeta}^{B_2}(\boldsymbol{\Phi}, \theta_1, \theta_2)\right|$ over the same angular grid as \Cref{fig:scf_patch} for the array operating at $B_2 Hz$ and using a linear operator that is (a) randomly drawn from a zero-mean complex normal distribution and (b) optimized using \gls*{sgd}. We see in (b) a reduction in the angular side-lobes as compared to \Cref{fig:scf_patch_nb}.}
    \label{fig:results}
\end{figure}
\subsection{Results} \label{sec:results}
\begin{figure}[h]
    \centering
    \includegraphics[width=\linewidth]{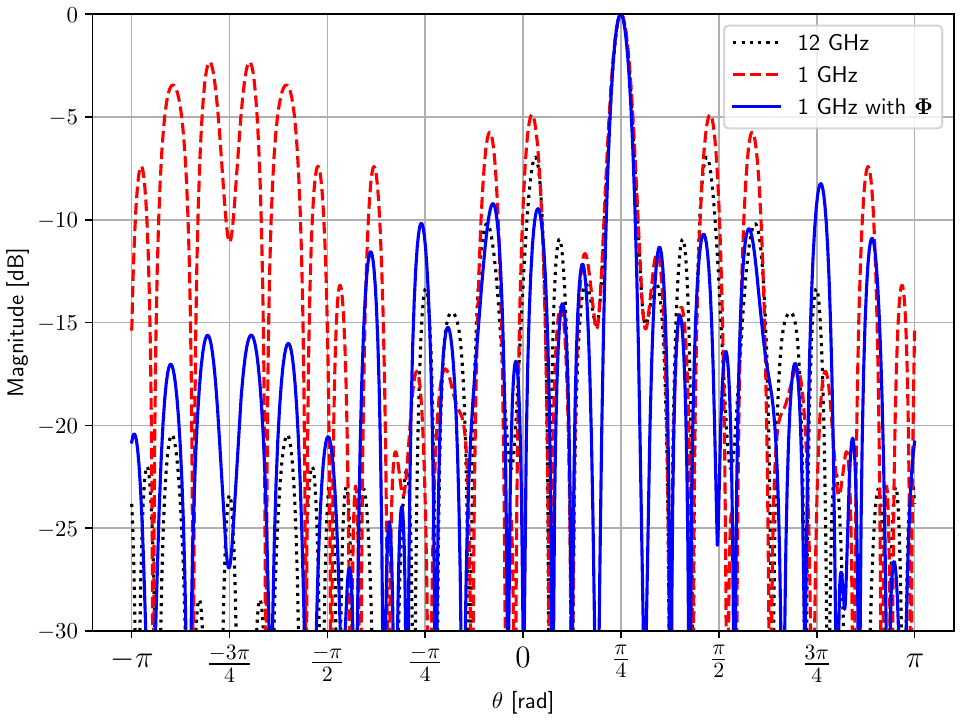}
    \caption{Magnitude of the correlation functions \eqref{eq:corr_func}, \eqref{eq:corr_func_phi} for a source located at azimuth angle $\pi/4$ for the system operating at \SI{\chaserbw}{\giga\hertz} bandwidth before (dashed) and after (solid) using the linear operator $\Phi$. Also visualized is the correlation function for the large bandwidth system (dotted).}
    \label{fig:fwd_atoms_corr}
\end{figure}

In \Cref{fig:results} we see the reduction of angular side-lobes in the effective \gls*{scf} attained when making use of linear operator $\Phi$ at the frequency $B_2 \ \SI{}{\hertz}$. It can be seen in \Cref{fig:scf_patch_mixed} that the design procedure of \Cref{sec:extension_op_design} outputs an operator $\boldsymbol{\Phi}$ that significantly reduces the off-diagonal elements in the effective \gls*{scf} as compared to the randomly drawn operator, illustrated in \Cref{fig:scf_patch_random}. Similarly, in \Cref{fig:fwd_atoms_corr} we see the reduction in angular side-lobes achievable in the correlation function of \eqref{eq:corr_func} for a single propagating path when utilizing the designed operator $\Phi$. 

\section{Conclusions}
The results in \Cref{sec:results} illustrate one method by which angular side-lobe reduction is achievable for arrays that have large inter-element distances. This suggests that the constraints imposed on the fabrication process of antenna arrays operating at the \gls*{mmwave}, sub-\SI{}{\tera\hertz} and \SI{}{\tera\hertz} can be relaxed by utilizing either a sufficient measurement bandwidth or effective post-processing that allows for a lower bandwidth.

Further investigation of the performance of the approach in \Cref{sec:reducing_angular_side} for scenarios with multi-path propagation is required. Initial steps in this direction have been taken and will be published subsequently.

\printbibliography

\end{document}

%% file: my_macros.tex
\newcommand{\R}{{\mathbb R}}
\newcommand{\C}{{\mathbb C}}

\def\targetbw{12}
\def\chaserbw{1}
\def\carrier{33}
\def\numElems{8}
\def\lamdaFactor{3}
\def\numFreq{32}

\def\betaOne{0.3}
\def\betaTwo{0.999}
\def\alphaLearn{0.001}

%% file: abbrevs.tex
\newacronym{aoa}{AoA}{Angle-of-Arrival}
\newacronym{aod}{AoD}{Angle-of-Departure}
\newacronym{adam}{ADAM}{ADAM}
\newacronym{cs}{CS}{Compressed Sensing}

\newacronym{eadf}{EADF}{Effective Aperture Distribution Function}

\newacronym{icas}{ICAS}{Integrated Communications and Sensing}

\newacronym{ir}{IR}{Impulse Response}
\newacronym{ml}{ML}{Maximum-Likelihood}
\newacronym{mmwave}{mmWave}{Millimeter-wave}

\newacronym{scf}{SCF}{Spatial Correlation Function}
\newacronym{sgd}{SGD}{Stochastic Gradient Descent}